\documentclass[aps,showpacs,preprintnumbers,amsmath, amssymb]{revtex4}

\usepackage{float}
\usepackage{graphics,epsfig}
\usepackage{graphicx}
\usepackage{dcolumn}
\usepackage{bm}

\begin{document}
\baselineskip=0.8 cm
\title{{\bf Holographic entanglement entropy in superconductor phase transition with dark matter sector}}
\author{Yan Peng$^{1}$\footnote{yanpengphy@163.com}}
\affiliation{\\$^{1}$ School of Mathematics and Computer Science, Shaanxi University of Technology, Hanzhong, Shaanxi 723000, P. R. China  }

\vspace*{0.2cm}
\begin{abstract}
\baselineskip=0.6 cm
\begin{center}
{\bf Abstract}
\end{center}

In this paper, we investigate the holographic phase transition with dark matter sector in the AdS black hole
background away from the probe limit.
We disclose the properties of phases mostly from
the holographic topological entanglement entropy of the system.
We find the entanglement entropy is a good probe to the critical
temperature and the order of the phase transition in the general model.
The behaviors of entanglement entropy at large strip size suggest that the area law still
holds when including dark matter sector.
We also conclude that the holographic topological entanglement entropy is useful in
detecting the stability of the phase transitions.
Furthermore, we derive the complete diagram of the effects of coupled parameters on the critical
temperature through the entanglement entropy and analytical methods.

\end{abstract}

\maketitle
\newpage
\vspace*{0.2cm}

\section{Introduction}

The AdS/CFT correspondence provides us a powerful approach to holographically study
strongly interacting low energy physics
in condensed matter systems. According to this correspondence,
the d dimensional strongly interacting theories on the boundary are
dual to the d+1 dimensional weakly coupled gravity theories in the bulk \cite{Maldacena,S.S.Gubser-1,E.Witten}.
The most simple holographic superconductor model dual to gravity theories is constructed by applying a scalar field and
a Maxwell field coupled in an AdS black hole background \cite{S.A. Hartnoll,C.P. Herzog,G.T. Horowitz-1}.
Since then, a lot of more complete holographic superconductor models were also taken into account,
such as the holographic superconductor models in Einstein-Gauss-Bonnet gravity,
Horava-Lifshitz gravity, non-linear electrodynamics gravity and so on.
These typical examples have attracted considerable interest
for their potential applications to the condensed matter physics,
see \cite{R}-\cite{MB}.

Recently, a gravity theory with dark matter sector was proposed in \cite{HD,TA}.
This new gravity theory was also considered in holographic superconductor models,
which was constructed with a scalar field, a Maxwell field and another
additional U(1)-gauge field corresponding to the dark matter one \cite{LN-1,LN-2}.
With Sturm-Liouville eigenvalue and matching semi-analytical methods,
it has been disclosed in \cite{LN-1} that the dark matter sector can bring rich physics in the new holographic model.
Very surprisingly, this new model also allow superconducting solutions corresponding to retrograde condensation.
In order to further study the effects of the dark matter sector on the holographic phase transition from other aspects of the scalar operator, the free energy and so on, we will have to turn to the numerical methods.

The papers \cite{S-1,S-2}
have shown us a novel way to calculate the holographic entanglement entropy of
a strongly interacting system from a weakly coupled gravity dual according to the AdS/CFT correspondence.
In this way, the holographic entanglement entropy has recently been
applied to study the properties of phase transitions in various holographic models
\cite{NishiokaJHEP}-\cite{Yan Peng-2}.
The entanglement entropy representing the degrees of freedom of the systems turns out to be a good probe to
investigate the critical temperature and the order of the holographic phase transition.
It was argued that the discontinuous slops imply the second order phase transition
and the jump of the holographic entanglement entropy corresponds
to the first order phase transition.
It is meaningful to examine whether the holographic entanglement entropy approach is still useful in the holographic
superconductor model with dark matter sector.
From the other aspect, it was found that the entanglement entropy and thermal entropy behave qualitatively the same
for large width strip \cite{BS,Cai-4}.
In accordance with the area law, it was found that the holographic entanglement entropy goes linearly for large strip \cite{T-6}.
Then it is expected that we can test the area law with dark matter sector from the entanglement entropy side.
At last, the retrograde condensation phenomenon was observed in \cite{LN-1}, which usually corresponds to
unstable solutions \cite{JD,FDJ}. We will also try to disclose the stability
of the retrograde condensation solutions through the holographic entanglement entropy method.

The next sections are organized as follows. In section II, we review the construction of the holographic
superconductor model with dark matter sector in the four dimensional AdS black hole spacetime beyond the probe limit.
In section III, we study the properties of the holographic phase transitions by
examining in detail the behaviors of the holographic entanglement entropy.
We also give some analytical understanding of the
phase transition properties.
We summarize our main results in the last section.

\section{Equations of motion and boundary conditions}

The holographic superconductor model with dark matter sector is constructed by a scalar field and two gauge fields
coupled in the AdS black hole background. The generalized Lagrange density of 4-dimensional spacetime with dark matter sector reads \cite{LN-1}:
\begin{eqnarray}\label{lagrange-1}
\mathcal{L}=R+\frac{6}{L^{2}}-\gamma[\frac{1}{4}F^{MN}F_{MN}+|\nabla_{M}\psi-iA_{M}|^{2}+m^{2}\psi^{2}+\frac{1}{4}B^{MN}B_{MN}
+\frac{\alpha}{4}B^{MN}F_{MN}],
\end{eqnarray}
where $\psi(r)$ is a complex scalar field with mass $m$.
$A_{M}$ stands for the ordinary Maxwell field and $B_{M}$ is the additional $U(1)$ gauge field representing the dark matter one.
 $-3/L^{2}$ is the negative cosmological constant, where $L$ is the AdS radius
which will be scaled unity in our calculation.
$\gamma$ describes the backreaction of matter fields on the background.
When $\gamma\rightarrow 0$, we return to the holographic model in the probe limit \cite{Y. Brihaye}.
$\alpha$ is the coupling parameter between the two gauge fields,
which should be small and on the order of $10^{-3}$ according to present astronomical observation \cite{NNL}.
In fact, when studying holographic superconductors, one is normally interested in the
CFT on the boundary. It is not an issue whether the dual bulk
gravitational theory is realistic or not.
So in this paper, we study $\alpha$ varying in large range $\alpha\in[0,2.5]$.

The Einstein equations for the system can be written in the form
\begin{eqnarray}\label{BHpsi}
R_{MN}-\frac{1}{2}g_{MN}R-3g_{MN}=\frac{1}{2}\gamma \tilde{T}_{MN},
\end{eqnarray}
where $\tilde{T}_{MN}$ is the energy-momentum tensor expressed as
\begin{equation}
\begin{aligned}
\tilde{ T}_{MN}=F_{M\beta}F^{\beta}_{N}+B_{M\beta}B^{\beta}_{N}+\alpha B_{M\beta}F^{\beta}_{N}+2\nabla_{M}\psi\nabla_{N}\psi
+2A_{M}A_{N}\psi^{2}~~~~~~~~~~~~~~~~~~~~~~~~~~~\\
+g_{MN}(-\frac{1}{4}F_{MN}F^{MN}-\frac{1}{4}B_{MN}B^{MN}
-\frac{\alpha}{4}F_{MN}B^{MN}-\nabla_{M}\psi\nabla_{M}\psi-A_{M}A^{M}\psi^{2}-m^{2}\psi^{2}).
 \end{aligned}
\end{equation}

With the variation of the  matter fields, we get the corresponding equations of motion:
\begin{eqnarray}\label{BHChi}
\nabla_{M}F^{MN}-2A^{N}\psi^{2}+\frac{\alpha}{2}\nabla_{M}B^{MN}=0,
\end{eqnarray}
\begin{eqnarray}\label{BHChi}
\nabla_{M}\nabla^{M}\psi-A_{M}A^{M}\psi-m^{2}\psi=0,
\end{eqnarray}
\begin{eqnarray}\label{BHChi}
\nabla_{M}B^{MN}+\frac{\alpha}{2}\nabla_{M}F^{MN}=0.
\end{eqnarray}

Putting (6) into (4), we get the equation
\begin{eqnarray}\label{BHChi}
\nabla_{M}F^{MN}-\frac{2\psi^{2}A^{N}}{\tilde{\alpha}}=0,
\end{eqnarray}
where $\tilde{\alpha}=1-\frac{\alpha^{2}}{4}$.

Putting (7) into (6), we arrive at
\begin{eqnarray}\label{BHChi}
\nabla_{M}B^{MN}+\frac{\alpha\psi^{2}A^{N}}{\tilde{\alpha}}=0.
\end{eqnarray}

In this work, we simply take the metric solutions and other matter fields in the forms:
\begin{eqnarray}\label{AdSBH}
ds^{2}&=&-g(r)e^{-\chi(r)}dt^{2}+\frac{dr^{2}}{g(r)}+r^{2}(dx^{2}+dy^{2}),
\end{eqnarray}
\begin{eqnarray}\label{symmetryBH}
A=\phi(r)dt,~~~~~~~~B=\eta(r)dt,~~~~~~~\psi=\psi(r).
\end{eqnarray}
Then the Hawking temperature of the black hole is expressed as
\begin{eqnarray}\label{HawkingT}
T=\frac{g'(r_{+})e^{-\chi(r_{+})/2}}{4\pi},
\end{eqnarray}
where $r_{+}$ is the horizon of the black hole satisfying $g(r_{+})=0$.
We also need $\chi(r\rightarrow\infty)=0$ to recover the AdS boundary.

From above assumptions, we can obtain the equations of motion as:
\begin{eqnarray}\label{BHpsi}
\psi''+\left(\frac{2}{r}-\frac{\chi'}{2}+\frac{g'}{g}\right)\psi'-\frac{m^{2}}{g}\psi+\frac{1}{g^{2}}e^{\chi}\phi^{2}\psi=0,
\end{eqnarray}
\begin{eqnarray}\label{BHphi}
\phi''+\left(\frac{2}{r}+\frac{\chi'}{2}\right)\phi'-\frac{2\psi^{2}}{\tilde{\alpha}g}\phi=0,
\end{eqnarray}
\begin{eqnarray}\label{BHphi}
\eta''+\left(\frac{2}{r}+\frac{\chi'}{2}\right)\eta'+\frac{\alpha}{\tilde{\alpha}}\frac{\psi^{2}}{g}\phi=0,
\end{eqnarray}
\begin{eqnarray}\label{BHg}
g'-\left(\frac{3r}{L^{2}}-\frac{g}{r}\right)+
\kappa^{2} r\left[2g\psi'^{2}+e^{\chi}(\phi'^{2}+\eta'^{2}+\alpha\phi'\eta')+2m^{2}\psi^{2}+\frac{2e^{\chi}\phi^{2}\psi^{2}}{g}\right]=0,
\end{eqnarray}
\begin{eqnarray}\label{BHChi}
\chi'+4\kappa^{2}\left[r\psi'^{2}+\frac{r}{g^{2}}e^{\chi}\phi^{2}\psi^{2}\right]=0,
\end{eqnarray}

Where $\kappa^{2}=\frac{\gamma}{4}$.
When a hairy black hole with $\psi \neq 0$ appears,
we have to solve these equations
by numerical methods.
Near the AdS boundary $(r\rightarrow \infty)$, the asymptotic behaviors of the solutions are
\begin{eqnarray}\label{InfBH}
\psi\rightarrow\frac{\psi_{-}}{r^{\lambda_{-}}}+\frac{\psi_{+}}{r^{\lambda_{+}}},~~~\
\phi\rightarrow \mu-\frac{\rho}{r},~~~ \eta\rightarrow\xi-\frac{\varpi}{r},~~~ f\rightarrow \frac{r^{2}}{L^{2}},~~~ \chi\rightarrow 0,\ \
\end{eqnarray}
with $\lambda_{\pm}=(3\pm\sqrt{9+4m^{2}})/2$, where $\mu$ and $\rho$ can be interpreted as
the chemical potential and charge density in the dual theory
respectively. The other two operators $\xi$ and $\varpi$ are dual to
the U(1) gauge field $\eta(r)$.

At the horizon of the black hole, we can impose proper boundary conditions as:
\begin{eqnarray}\label{InfBH}
\psi(r)=\psi_{0}+\psi_{1}(r-r_{+})+\cdots,~~~~~~~~~~
\end{eqnarray}
\begin{eqnarray}\label{InfBH}
\phi(r)=\phi_{0}(r-r_{+})+\phi_{1}(r-r_{+})^{2}+\cdots,
\end{eqnarray}
\begin{eqnarray}\label{InfBH}
\eta(r)=\eta_{0}(r-r_{+})+\eta_{1}(r-r_{+})^{2}+\cdots,
\end{eqnarray}
\begin{eqnarray}\label{InfBH}
g(r)=g_{0}(r-r_{+})+g_{1}(r-r_{+})^{2}+\cdots,
\end{eqnarray}
\begin{eqnarray}\label{InfBH}
\chi(r)=\chi_{0}+\chi_{1}(r-r_{+})+\cdots,~~~~~~~~~~
\end{eqnarray}
where the dots denote higher order terms.
Putting these Taylor expansions into equations,
we are left with five independent parameters $r_{+}$,
$\psi_{0}$, $\phi_{0}$, $\eta_{0}$ and $\chi_{0}$ at the horizon. The scaling symmetry
\begin{eqnarray}\label{symmetryBH}
r \rightarrow ar,~~~~~~~~t\rightarrow
~at,~~~~~~~\phi\rightarrow
a\phi,~~~~~~~\eta\rightarrow
a\eta,~~~~~~g\rightarrow\ a^{2} g,
\end{eqnarray}
can be used to set $r_{+}=1$.

Choosing $m^{2}=-2>-\frac{9}{4}$ above the BF bound \cite{P. Breitenlohner}, the second mode $\psi_{+}$ is always normalizable.
To get a stable theory, we will fix $\psi_{-}=0$ and use the operator $\psi_{+}=<O_{+}>$ to describe the phase transition in the dual CFT.
For different values of $\psi_{0}$, we can rely on the parameters $\phi_{0}$, $\eta_{0}$ and $\chi_{0}$
to search for the solutions with the boundary conditions $\psi_{-}=0$, $\chi(r\rightarrow\infty)=0$ and $\frac{\xi}{\mu}$ fixed
as an constant.
We mention that the solutions go back to the case without dark matter sector when $\frac{\xi}{\mu}\rightarrow0$.
In the following discussion, we will divide the phases into two cases: $1+\alpha\frac{\xi}{\mu}+\frac{\xi^{2}}{\mu^{2}}>0$
corresponds to the thermodynamically stable solutions and $1+\alpha\frac{\xi}{\mu}+\frac{\xi^{2}}{\mu^{2}}<0$ is the
thermodynamically unstable phase transition.

In the case of $\psi=0$, we get the exact analytic solutions in normal phase,
a Reissner-Nordstrom-AdS black hole, which is given by
\begin{eqnarray}\label{}
g=\frac{r^{2}}{L^{2}}-\frac{2\tilde{M}}{r}+\frac{\kappa^{2}(\rho^{2}+\varpi^{2}+\alpha\rho\varpi)}{r^{2}} ,~~\chi=0,~~~\phi=\rho(\frac{1}{r_{+}}-\frac{1}{r}),~~\eta=\varpi(\frac{1}{r_{+}}-\frac{1}{r}),
\end{eqnarray}
where $\tilde{M}$ can be interpreted as the mass of the black hole.

\section{holographic phase transitions in AdS black hole background}

\subsection{The stable phases with $1+\alpha\frac{\xi}{\mu}+\frac{\xi^{2}}{\mu^{2}}>0$}

In this part, we focus on the holographic entanglement entropy(HEE) of the phase transition satisfying $1+\alpha\frac{\xi}{\mu}+\frac{\xi^{2}}{\mu^{2}}>0$.
The authors in Refs. \cite{S-1,S-2} have provided a method to compute
the entanglement entropy of conformal field theories (CFTs) from the gravity side.
We consider a belt geometry $\tilde{A}$ with a finite width along the x direction and infinitely extending in y
direction as: $-\frac{l}{2}\leqslant x\leqslant\frac{l}{2},~0\leq y \leq \tilde{L}$,
where $l$ is defined as the size of region $\tilde{A}$, and
$\tilde{L}$ is a regulator which is set to infinity. Minimizing
the area of hypersurface $\gamma_{\tilde{A}}$ whose boundary is the
same as the stripe $\tilde{A}$, we can deduce the entanglement entropy for $\tilde{A}$ as \cite{T-6}
\begin{eqnarray}\label{EEntropyBH}
S=\int^{z_{*}}_{\varepsilon}dz\frac{z_{*}^{2}}{z^{2}}\frac{1}{\sqrt{(z^{4}_{*}-z^{4})z^{2}g(z)}}-\frac{1}{\varepsilon},
\end{eqnarray}
with
\begin{eqnarray}\label{Length}
\frac{l}{2}=\int^{z_{*}}_{\varepsilon}dz\frac{z^{2}}{\sqrt{(z^{4}_{*}-z^{4})z^{2}g(z)}},
\end{eqnarray}
where $z_{*}$ satisfies the condition $\frac{dz}{dx}|_{z_{*}}=0$
with $z=\frac{1}{r}$ and $r=\frac{1}{\varepsilon}$ is the UV cutoff.

\begin{figure}[h]
\includegraphics[width=180pt]{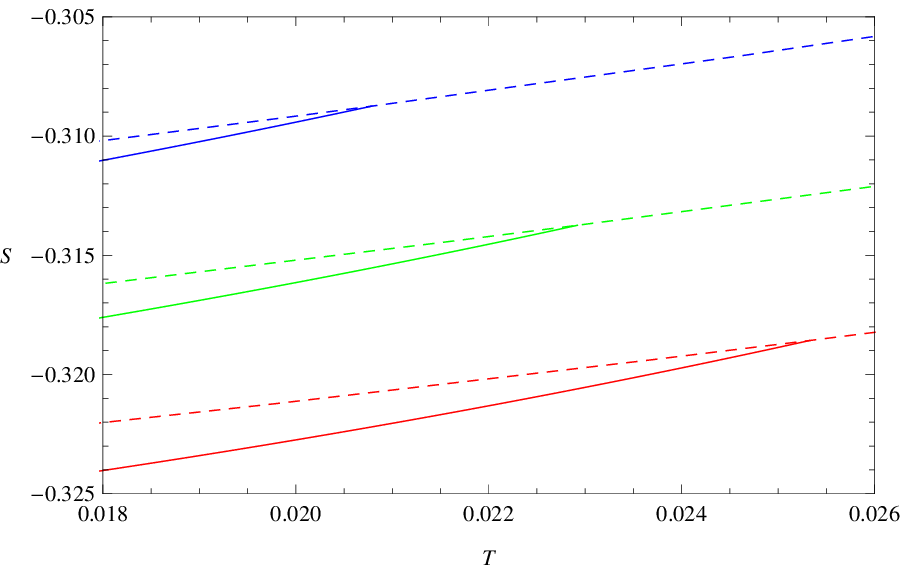}\
\includegraphics[width=180pt]{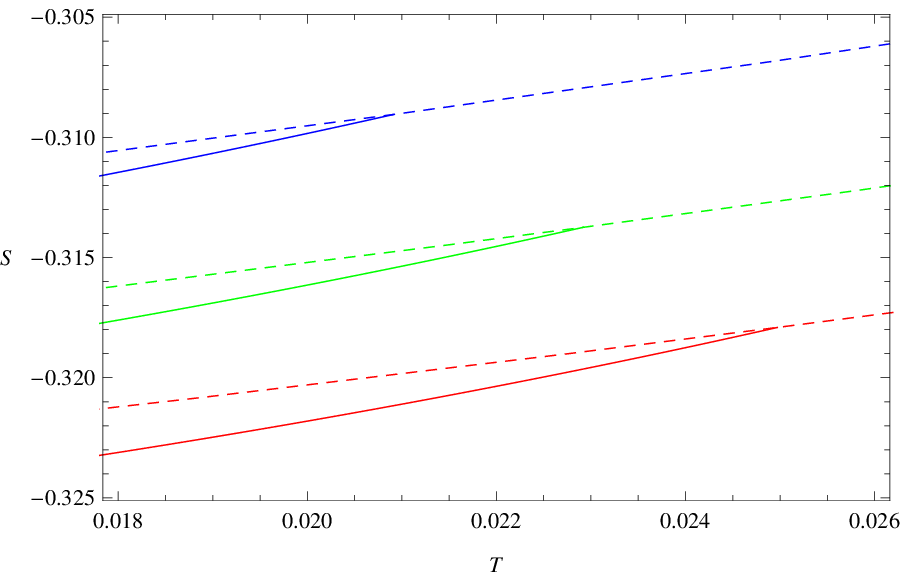}\
\caption{\label{EEntropySoliton} (Color online) The holographic entanglement
entropy as a function of the temperature $T$ for
$\mu=1$, $\kappa^{2}=0.1$, $m^{2}=-2$ and $l=2$. The dashed lines in each panel corresponds to the entanglement
entropy of a pure AdS black hole and the solid lines correspond to cases of superconducting solutions.
In the left panel, we have choose $\frac{\xi}{\mu}=1$ and various $\alpha$: $\alpha=0$ with $T_{c}= 0.02533$ (Red); $\alpha=0.25$ with $T_{c}= 0.02292$ (Green); $\alpha=0.5$ with $T_{c}= 0.02079$ (Blue).
The right panel correspond to $\alpha=0.25$ and various $\frac{\xi}{\mu}$: $\frac{\xi}{\mu}=0.9$ with $T_{c}=0.02497$ (Red);
$\frac{\xi}{\mu}=1.0$ with $T_{c}=0.02292$ (Green); $\frac{\xi}{\mu}=1.1$ with $T_{c}=0.02091$ (Blue).}
\end{figure}

We present the holographic entanglement entropy as a function of the temperature $T$ in Fig. 1 with $\mu=1$, $\kappa^{2}=0.1$, $m^{2}=-2$, $\frac{\xi}{\mu}>0$
and $l=2$.
The left panel is the cases of $\mu=1$, $\kappa^{2}=0.1$, $m^{2}=-2$, $\frac{\xi}{\mu}=1$, $l=2$ with various $\alpha$
and right panel shows the cases
of $\mu=1$, $\kappa^{2}=0.1$, $m^{2}=-2$, $\alpha=0.25$, $l=2$ with various $\frac{\xi}{\mu}>0$.
It can be easily seen from the pictures that when the parameters fixed, the entanglement entropy decreases monotonously as we choose a smaller  Harking temperature T. It means there is a reduction in the number of degrees
of freedom due to the condensate generated in the phase transitions \cite{Cai-4}.

Decreasing the temperature, we should determine the physical curve by always
choosing the point of lowest entropy at a given T \cite{T-6}.
For each set of parameters, we find a threshold temperature $T_{c}$, below which the hairy black hole appears.
We also obtain approximate formulas for the holographic entanglement entropy
of the normal state ($S_{BH}$) and superconducting state ($S_{SC}$) in the case of $\mu=1$, $\kappa^{2}=0.1$, $m^{2}=-2$, $l=2$, $\alpha=0.5$ and $\frac{\xi}{\mu}=1$:
\begin{eqnarray}\label{BHpsi}
S_{BH}\thickapprox -0.309+0.557(T-T_{c})+3.434(T-T_{c})^2,\\
S_{SC}\thickapprox -0.309+0.862(T-T_{c})+17.273(T-T_{c})^2.
\end{eqnarray}
These formulas suggest that $S_{BH}|_{T=T_{c}}=S_{SC}|_{T=T_{c}}$, $\frac{\partial S_{BH}}{\partial T}|_{T=T_{c}}\neq\frac{\partial S_{SC}}{\partial T}|_{T=T_{c}}$ and the jump of the slop of the entanglement entropy signal that
some kind of new degrees of freedom like the Cooper pair would emerge in the new phase.

In order to show there is a phase transition, we calculate the
free energy of the system with $\mu=1$, $\kappa^{2}=0.1$, $m^{2}=-2$, $\alpha=0.5$ and $\frac{\xi}{\mu}=1$ in Fig. 2.
The free energy of superconducting state lies below the free energy of the normal state suggesting that the superconducting solutions are thermodynamically stable.
More calculations show that there are thermodynamically stable solutions for all $1+\alpha\frac{\xi}{\mu}+\frac{\xi^{2}}{\mu^{2}}>0$.

\begin{figure}[h]
\includegraphics[width=180pt]{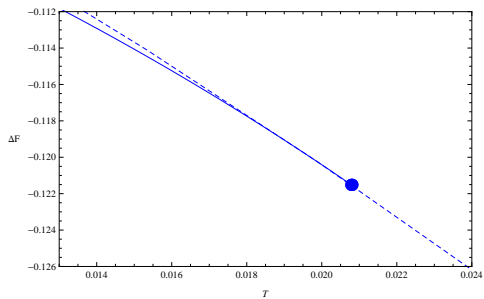}\
\caption{\label{EEntropySoliton} (Color online) The free energy in cases of $\mu=1$, $m^{2}=-2$, $\kappa^{2}=0.1$, $\alpha=0.5$ and $\frac{\xi}{\mu}=1$.
The solid blue line shows the free energy of the superconducting phase. The dashed line corresponds to the free energy of normal phases.
The solid blue point corresponds to the phase transition temperature $T=0.02079$.
}
\end{figure}

As shown with the solid blue point in Fig. 2,
we find a critical phase transition temperature $T=0.02079$, which is
equal to the threshold temperature $T_{c}= 0.02079$ obtained from the behaviors of holographic entanglement entropy.
That means the holographic entanglement entropy can be used to search for
the critical phase transition temperature.
With fitting methods, we obtain approximate formulas
for the free energy of normal state ($F_{BH}$) and superconducting state ($F_{SC}$) that:
\begin{eqnarray}\label{BHpsi}
F_{BH}\thickapprox -0.122-1.417(T-T_{c})-10.136(T-T_{c})^2-38.412(T-T_{c})^3,\\
F_{SC}\thickapprox -0.122-1.417(T-T_{c})-23.573(T-T_{c})^2-190.175(T-T_{c})^3.
\end{eqnarray}
The front formulas suggest that: $F_{BH}|_{T=T_{c}}=F_{SC}|_{T=T_{c}}$, $\frac{\partial F_{BH}}{\partial T}|_{T=T_{c}}=\frac{\partial F_{SC}}{\partial T}|_{T=T_{c}}$, $\frac{\partial^{2} F_{BH}}{\partial T^{2}}|_{T=T_{c}}\neq\frac{\partial^{2} F_{SC}}{\partial T^{2}}|_{T=T_{c}}$.
In other words, the curve representing the physical phases with the lowest free energy is smooth at the critical temperature $T_{c}$.
We conclude that the jump of the slope of holographic entanglement entropy
corresponds to second order phase transition in the general holographic superconductor model with dark matter.

\begin{figure}[h]
\includegraphics[width=180pt]{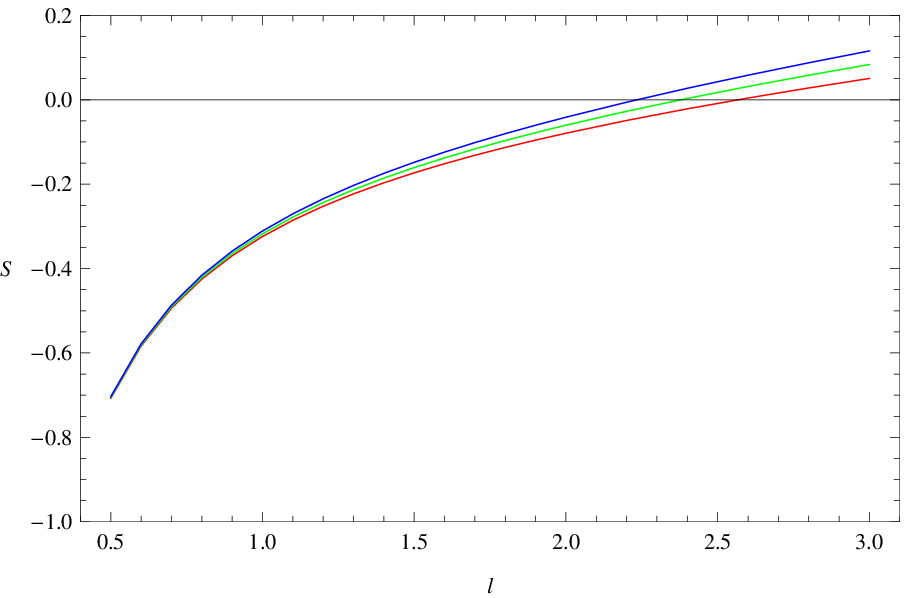}\
\includegraphics[width=180pt]{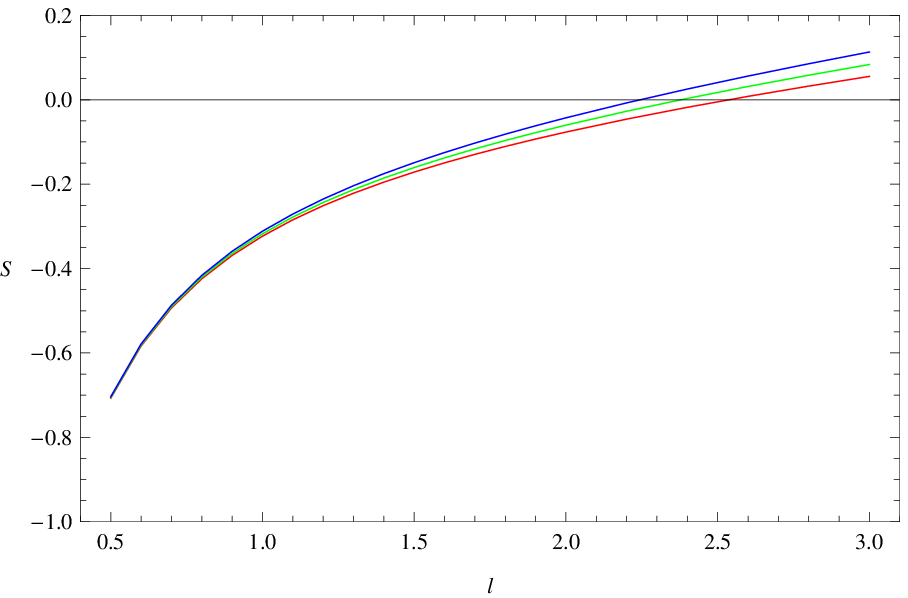}\
\caption{\label{EEntropySoliton} (Color online) The holographic entanglement
entropy as a function of the belt $l$ with
$\mu=1$, $\kappa^{2}=0.1$, $m^{2}=-2$ and $T=0.018$. The solid lines correspond to cases of superconducting solutions.
In the left panel, we have choose $\frac{\xi}{\mu}=1$ and various $\alpha$ from bottom to top: $\alpha=0$(Red), $\alpha=0.25$(Green), $\alpha=0.5$(Blue).
The right panel from bottom to top corresponds to $\alpha=0.25$ and various $\frac{\xi}{\mu}$: $\frac{\xi}{\mu}=0.9$(Red), $\frac{\xi}{\mu}=1.0$(Green), $\frac{\xi}{\mu}=1.1$(Blue).}
\end{figure}

We also show the behaviors of the holographic entanglement
entropy S with respect to the strip width $l$ at a fixed temperature T=0.018 below the phase transition temperature $T_{c}$ in Fig. 3.
The solid colour lines denote the holographic entanglement
entropy for superconducting phase. We see that for each line,
S increases monotonically from a negative value to a
positive value as we increase $l$.
Choosing a larger strip size $l$, the entanglement entropy becomes sensitive to $\alpha$ and $\frac{\xi}{\mu}$.
In all panels, the curves go linearly with $l$ for large $l$, which means the area law holds in
holographic models with dark matter sector \cite{T-6}.

\begin{figure}[h]
\includegraphics[width=180pt]{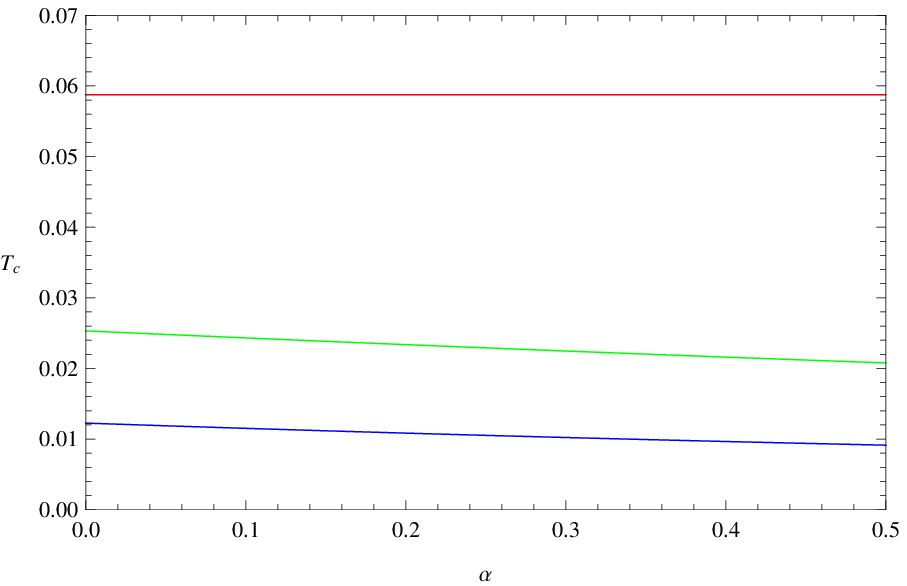}\
\includegraphics[width=180pt]{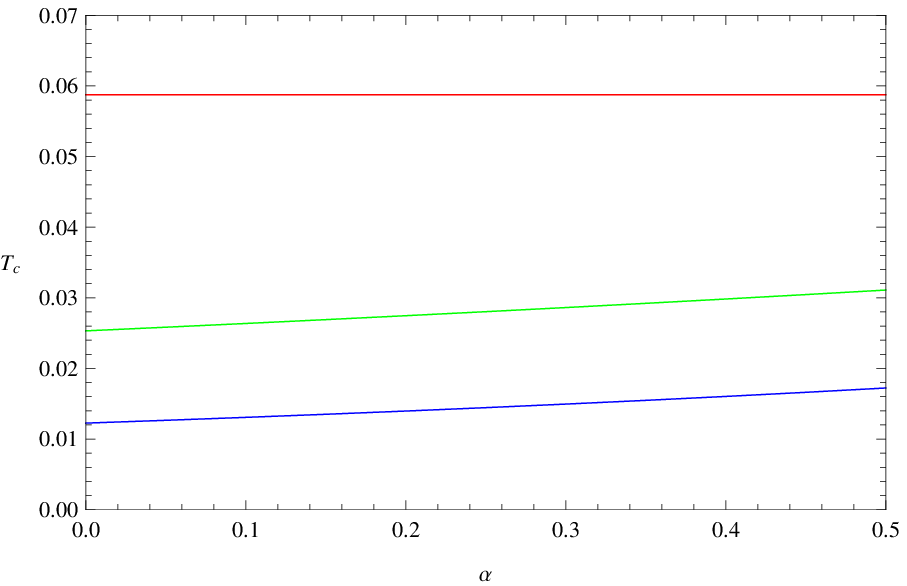}\
\caption{\label{EEntropySoliton} (Color online) The behaviors of the critical temperature $T_{c}$ with $\mu=1$, $m^{2}=-2$ and
the lines from top to bottom correspond to $\kappa^{2}=0$ (Red), $\kappa^{2}=0.1$ (Green), $\kappa^{2}=0.2$ (Blue).
The left panel represents the case of $T_{c}$ as a function of $\alpha$ with $\frac{\xi}{\mu}=1$ . The right panel shows the case of $T_{c}$ as a function of $\alpha$ with $\frac{\xi}{\mu}=-1$.}
\end{figure}

By applying the holographic entanglement entropy method, we go on to disclose in detail
behaviors of the critical temperature $T_{c}$.
We show the effects of $\alpha$ on the critical temperature $T_{c}$ in Fig. 4.
In the left panel, when we increase the value of $\alpha$ in cases of $\kappa^{2}>0$ and $\frac{\xi}{\mu}=1$,
$T_{c}$ decreases slowly, which means that the phase transition is more difficult to happen.
In contrast, in the right panel, with $\frac{\xi}{\mu}=-1$ and $\kappa^{2}>0$,
we find that $T_{c}$ increase as a function of $\alpha$ or the phase transition is more easier to happen.
$T_{c}$ is always an constant $T_{c}\equiv 0.05875$ for different values of $\alpha$ with $\kappa^{2}=0$ in both panels.

\begin{figure}[h]
\includegraphics[width=180pt]{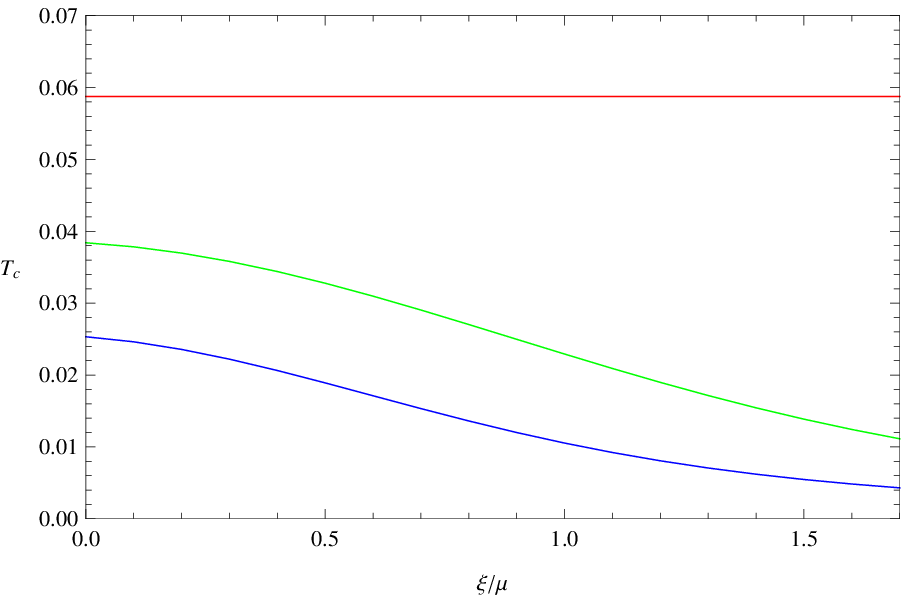}\
\includegraphics[width=180pt]{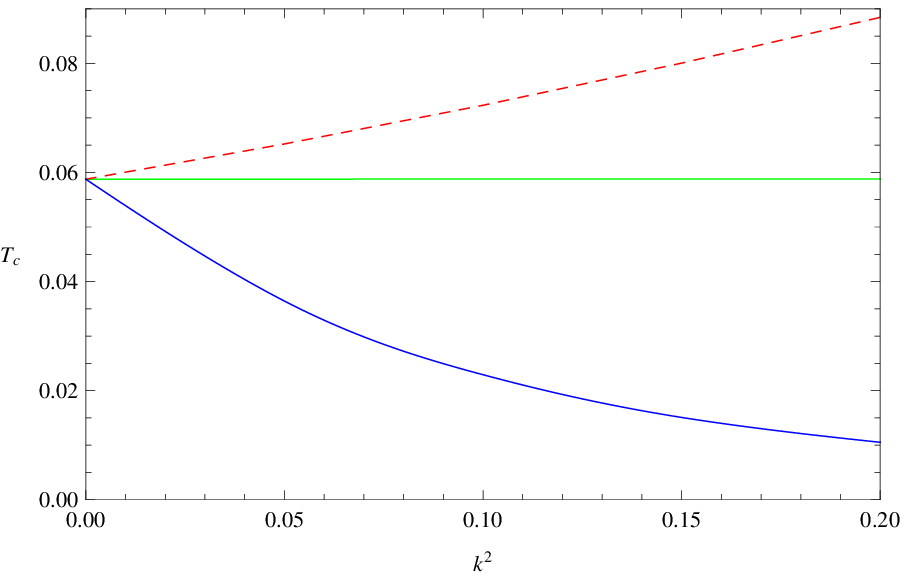}\
\caption{\label{EEntropySoliton} (Color online) The behaviors of the critical temperature $T_{c}$ with $\mu=1$, $m^{2}=-2$.
The lines in the left panel from top to bottom correspond to $\alpha=0.25$ and various $\kappa^{2}$: $\kappa^{2}=0$ (Red), $\kappa^{2}=0.1$ (Green), $\kappa^{2}=0.2$ (Blue).
The lines in the right panel from top to bottom represent the cases: (1) $\alpha=2.5$, $\frac{\xi}{\mu}=-1$ (Red);
(2) $\alpha\rightarrow 2$, $\frac{\xi}{\mu}=-1$ (Green); (3) $\alpha=0.25$, $\frac{\xi}{\mu}=1$ (Blue).}
\end{figure}

Now we turn to study how $\frac{\xi}{\mu}$ and $\kappa^{2}$ will affect the critical temperature $T_{c}$.
The curves in the left panel of Fig. 5 represent $T_{c}$ as a function of $\frac{\xi}{\mu}$ with $\mu=1$, $m^{2}=-2$, $\alpha=0.25$ and $\kappa^{2}$=0, 0.1 or 0.2.
It can be easily seen from the left panel that $T_{c}$ decreases as the ratio
becomes larger with $\kappa^{2}>0$.
We conclude that larger ratio $\frac{\xi}{\mu}>0$ makes it more difficult for the scalar field to condense
when considering the matter fields' backreaction
on the background.
Here, $T_{c}$ is also an constant $T_{c}\equiv 0.05875$ when neglecting backreaction.

In the right panel of Fig. 5, the bottom solid blue line shows that heavier backreaction corresponds to a smaller $T_{c}$
in the case of $\mu=1$, $m^{2}=-2$, $\alpha=0.25$ and $\frac{\xi}{\mu}=1$.
With more calculations, we conclude that larger $\kappa^{2}$ makes the scalar fields
more difficult to condense in all thermodynamically stable phases.
We observe an constant $T_{c}=0.05875$
in the case of $\mu=1$, $m^{2}=-2$, $\alpha\rightarrow 2$ and $\frac{\xi}{\mu}=-1$.
As a complete analysis, we also show behaviors of the critical temperature of thermodynamically unstable phases
in the case of $\mu=1$, $m^{2}=-2$, $\alpha=2.5$ and $\frac{\xi}{\mu}=-1$
satisfying $1+\alpha\frac{\xi}{\mu}+\frac{\xi^{2}}{\mu^{2}}<0$.
It is surprising that
a larger backreaction parameter leads to a larger $T_{c}$ for unstable solutions.
The former analytical results in \cite{LN-1} are included in the left panel of Fig. 4 and the right panel of Fig. 5.
So we obtain richer physics through holographic entanglement entropy approach.

\begin{figure}[h]
\includegraphics[width=180pt]{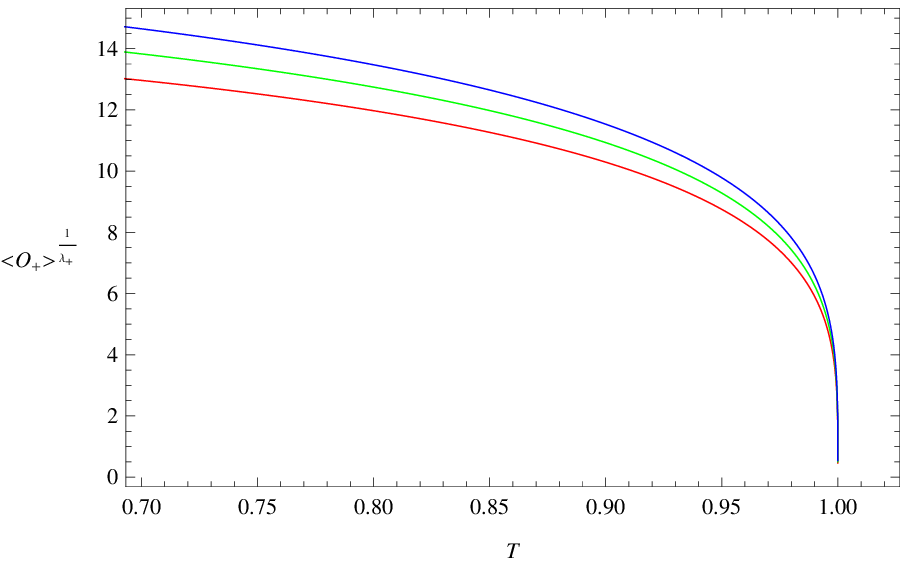}\
\includegraphics[width=180pt]{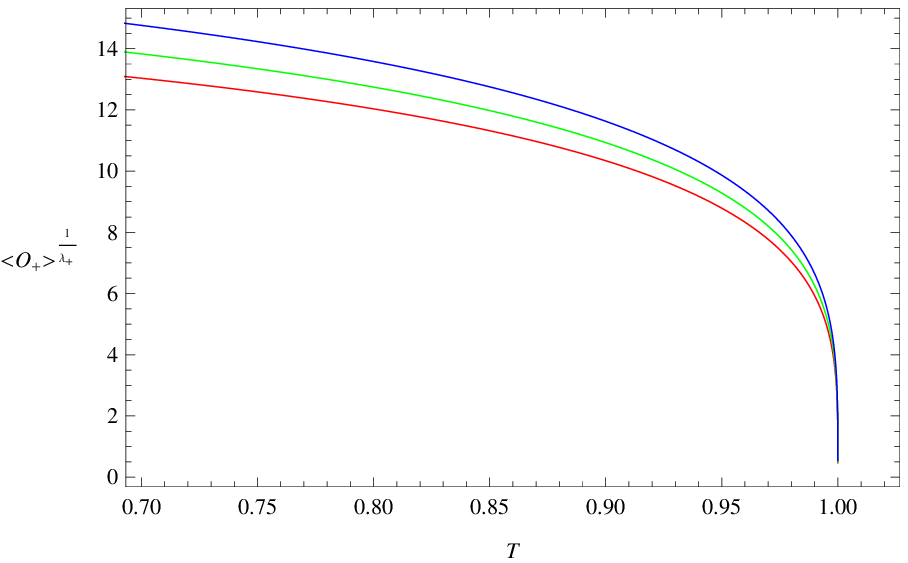}\
\caption{\label{EEntropySoliton} (Color online) The scalar operator as
a function of $T$ with $\mu=1$, $\kappa^{2}=0.1$, and $m^{2}=-2$. The left panel is with $\frac{\xi}{\mu}=1$
and various lines from bottom to top as: $\alpha=0$ (Red, $T_{c}=0.02533$), $\alpha=0.25$ (Green, $T_{c}=0.02292$) and $\alpha=0.5$ (Blue, $T_{c}=0.02079$).
The right panel is with $\alpha=0.25$
and various lines from bottom to top as: $\frac{\xi}{\mu}=0.9$ (Red, $T_{c}=0.02497$), $\frac{\xi}{\mu}=1.0$ (Green, $T_{c}=0.02292$) and $\frac{\xi}{\mu}=1.1$ (Blue, $T_{c}=0.02091$).}
\end{figure}

At last, we try to examine the correspondence between condensation gap and critical temperature \cite{Sean A. Hartnoll-3}.
We show the condensation
of the scalar operator $\langle O_{+}\rangle^{1/\lambda_{+}}$ with $T_{c}=1$, $\kappa^{2}=0.1$ and  $m^{2}=-2$ in Fig. 6.
In the left panel, we exhibit the condensation of the scalar operator
by choosing various $\alpha$ as: $\alpha=0$, $\alpha=0.25$, $\alpha=0.5$ and $\frac{\xi}{\mu}=1$.
The increase of the parameter $\alpha$
develops deeper condensation gap. From another aspect, the increase of $\alpha$ as:
$\alpha=0$, $\alpha=0.25$ and $\alpha=0.5$ corresponds to $T_{c}=0.02533$, $T_{c}=0.02292$ and $T_{c}=0.02079$ respectively.
In the right panel, we choose different ratios of $\frac{\xi}{\mu}$ and other parameters fixed to detect the behaviors of
phase transitions.
It is shown that larger ratios $\frac{\xi}{\mu}$ develop deeper condensation gap and the different
values of $\frac{\xi}{\mu}$: $\frac{\xi}{\mu}=0.9$, $\frac{\xi}{\mu}=1$ and $\frac{\xi}{\mu}=1.1$ correspond
to the critical temperatures $T_{c}=0.02497$, $T_{c}=0.02292$ and $T_{c}=0.02091$ respectively.
With various parameters $\alpha$ and $\frac{\xi}{\mu}$,
we arrive at the correspondence of a deeper condensation gap with a smaller $T_{c}$ in the holographic model with dark matter sector.

\subsection{The unstable phases with $1+\alpha\frac{\xi}{\mu}+\frac{\xi^{2}}{\mu^{2}}<0$}

In this part, we pay attention to the phase transitions satisfying $1+\alpha\frac{\xi}{\mu}+\frac{\xi^{2}}{\mu^{2}}<0$.
We study the example $\mu=1$, $m^{2}=-2$, $\kappa^{2}=0.1$, $\alpha=2.5$ and $\frac{\xi}{\mu}=-1$ in Fig. 7.
We show the holographic entanglement entropy in the left panel.
Choosing the phases with lowest entanglement entropy,
it is surprising that the condensed phase appears at high temperature $T>T_{c}=0.07231$,
which implies that this solution is unstable.
This novel behavior was referred as retrograde condensation \cite{LN-1,JD,FDJ}. We find that there are solutions of retrograde condensation phenomenon for all $1+\alpha\frac{\xi}{\mu}+\frac{\xi^{2}}{\mu^{2}}<0$.

\begin{figure}[h]
\includegraphics[width=180pt]{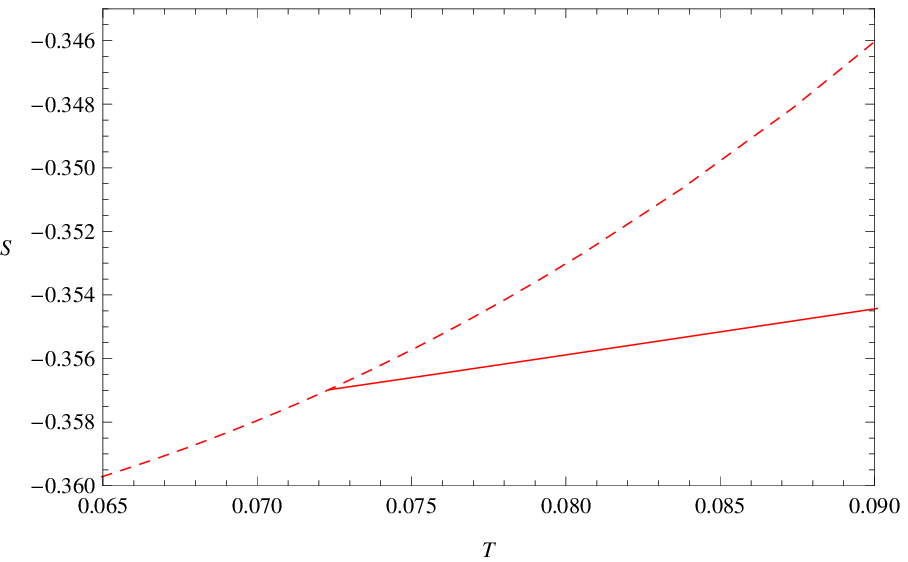}\
\includegraphics[width=180pt]{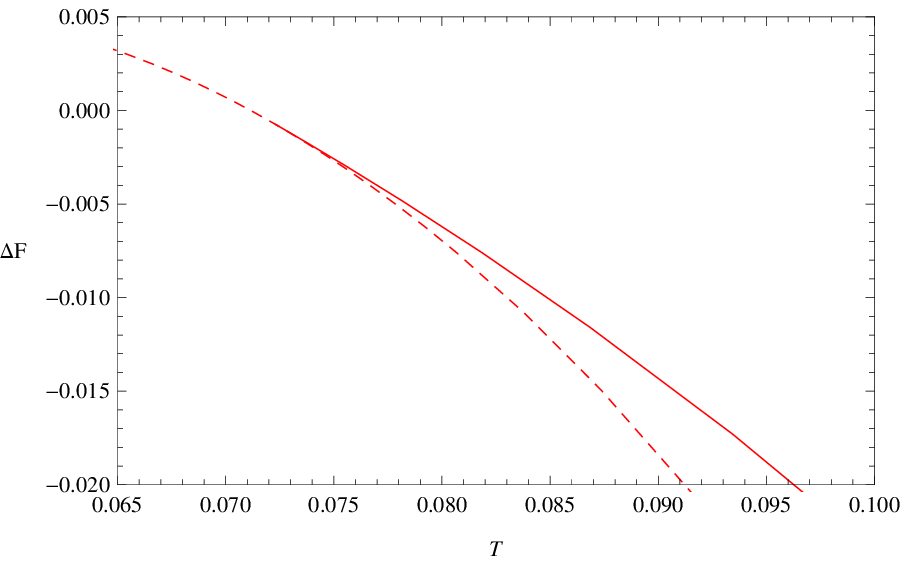}\
\caption{\label{EEntropySoliton} (Color online) The phase transition in the cases of $\mu=1$, $m^{2}=-2$, $\kappa^{2}=0.1$, $\alpha=2.5$ and $\frac{\xi}{\mu}=-1$. The left panel shows the behavior of holographic entanglement entropy with $l=2$.
The middle panel represents the free energy of the system. In both panels,
the solid red line corresponds to the superconducting phase and the dashed red line is with the normal phase.
}
\end{figure}

The free energy is powerful in studying the phase transitions.
We calculate the
free energy of the system in the right panel of Fig. 7.
It shows the free energy of
this hairy black hole is larger than the free energy of the black hole in normal phase.
Since the physical procedure corresponds to the phases with the lowest free energy,
we arrive at an conclusion that the retrograde condensation superconductor solutions are thermodynamically unstable
and the novel behaviors of the holographic entanglement entropy can be used to detect the thermodynamical stability of the phase transition.

\subsection{Analytical study of the condensation}

We would like to give an analytical understanding at the qualitative effects of $\alpha$, $\frac{\xi}{\mu}$ and $\kappa^{2}$ on the phase
transitions.
At the phase transition point, we have $\psi(r)=0$.
Then we can rewrite the equations of motion around $T_{c}$ in the form
\begin{eqnarray}\label{BHphi}
\phi''+\left(\frac{2}{r}+\frac{\chi'}{2}\right)\phi'=0,~~~~~\eta''+\left(\frac{2}{r}+\frac{\chi'}{2}\right)\eta'=0,
\end{eqnarray}
\begin{eqnarray}\label{BHg}
g'-\left(\frac{3r}{L^{2}}-\frac{g}{r}\right)+
\kappa^{2} r\left[e^{\chi}(\phi'^{2}+\eta'^{2}+\alpha\phi'\eta')\right]=0,~~~~\chi'=0.
\end{eqnarray}
According to the behaviors of the equations, we can assume the solutions $\eta(r)=\frac{\xi}{\mu}\phi(r)$, $\chi=0$ and deduce the equations for
$\phi(r)$ and $g(r)$ without dark matter term
\begin{eqnarray}\label{BHphi}
\phi''+\left(\frac{2}{r}+\frac{\chi'}{2}\right)\phi'=0,
\end{eqnarray}
\begin{eqnarray}\label{BHg}
g'-\left(\frac{3r}{L^{2}}-\frac{g}{r}\right)+
\tilde{\gamma} r\phi'^{2}=0,
\end{eqnarray}
where $\tilde{\gamma}=\kappa^{2} (1+\alpha\frac{\xi}{\mu}+\frac{\xi^{2}}{\mu^{2}})$.
We can taken $\tilde{\gamma}$
as the effective backreaction parameter to detect the changes of the critical temperature
on the idea that a larger effective backreaction parameter $\tilde{\gamma}$ corresponds
to a smaller critical temperature $T_{c}$ \cite{QB,Y. Brihaye}.

When $\kappa^{2}=0$, $\tilde{\gamma}$ is independent of $\alpha$ and $\frac{\xi}{\mu}$.
This is in accordance with the fact that $T_{c}$ keeps as an constant for different values of $\alpha$
and $\frac{\xi}{\mu}$ in cases of $\kappa^{2}=0$ in Fig. 4 and Fig. 5.
If we choose a larger $\alpha$ with fixed $\frac{\xi}{\mu}>0$ and $\kappa^{2}>0$, $\tilde{\gamma}$ increases
and the critical temperature $T_{c}$ becomes smaller.
With $\frac{\xi}{\mu}=-1<0$ and $\kappa^{2}>0$, a larger $\alpha$ leads to
a smaller $\tilde{\gamma}$ and a larger $T_{c}$.
We have now covered all the qualitative properties of Fig. 4.

In the left panel of Fig. 5,
a larger $\frac{\xi}{\mu}>0$ leads to a larger $\tilde{\gamma}$
and smaller $T_{c}$ when $\alpha>0$ and $\kappa^{2}>0$.
Now we try to see how $\kappa^{2}$ will affect the threshold temperature $T_{c}$.
In the case of $(1+\alpha\frac{\xi}{\mu}+\frac{\xi^{2}}{\mu^{2}})>0$,
a larger $\kappa^{2}$ corresponds to a larger $\tilde{\gamma}$ and a smaller $T_{c}$.
$\tilde{\gamma}$ is equal to zero for $(1+\alpha\frac{\xi}{\mu}+\frac{\mu^{2}}{\xi^{2}})=0$.
In cases that $(1+\alpha\frac{\xi}{\mu}+\frac{\xi^{2}}{\mu^{2}})<0$,
$\tilde{\gamma}$ becomes smaller and $T_{c}$ becomes larger when we choose a larger $\kappa^{2}$.
Accordingly, the three lines in the right panel of Fig. 5 from bottom to top
correspond to the decrease of $(1+\alpha\frac{\xi}{\mu}+\frac{\xi^{2}}{\mu^{2}})$
as: 2.25, 0 and -0.5 respectively.
We argue that negative effective backreaction parameters with thermodynamical instability
is a general properties in holographic phase transitions.

\section{Conclusions}

We studied the behaviors of holographic metal/superconductor phase
transitions in the presence of dark matter sector.
We tried to explore the properties of the phase transitions by analyzing the holographic
entanglement entropy of the system.
It was showed that the entanglement entropy can be used to search
the critical temperature and the jump of the slop of the holographic topological entanglement entropy
corresponds to a second order phase transitions when including the dark matter sector.
For larger strip size, the
behaviors of holographic topological entanglement entropy suggest that the area law still
holds with dark matter sector.
We also found that the entanglement entropy serves as a good probe to the stability of the phase transitions.
We obtained the thermodynamically stable conditions $1+\alpha\frac{\xi}{\mu}+\frac{\xi^{2}}{\mu^{2}}>0$ and our discussion are based on stable phases. In summary, we arrived at the conclusion that the holographic
entanglement entropy can be used to explore the rich physics in holographic superconductor models
with dark matter sector.

We also derived the complete diagram of effects of the parameters
$\alpha$, $\frac{\xi}{\mu}$ and $\kappa^{2}$ on the critical temperature $T_{c}$ through the entanglement entropy methods.
We found $\alpha$ and $\frac{\xi}{\mu}$ don't affect the critical temperature when neglecting the backreaction.
The larger positive parameters $\frac{\xi}{\mu}$ and $\kappa^{2}$ make the phase transition more difficult to happen.
When $\frac{\xi}{\mu}>0$ and $\kappa^{2}>0$, larger $\alpha$ makes the scalar fields harder
to condense and larger $\alpha$ makes the scalar hair easier to form in the case of $\frac{\xi}{\mu}<0$ and $\kappa^{2}>0$.
With various $\alpha$ and $\frac{\xi}{\mu}$, we also found that
a smaller critical temperature corresponds to a deeper condensation gap.
In all, we have obtained richer physics than the former analytical results in \cite{LN-1}.
By taking $\tilde{\gamma}=k^{2}(1+\alpha\frac{\xi}{\mu}+\frac{\xi^{2}}{\mu^{2}})$ as the effective backreaction parameter,
the qualitative properties can be obtained through the analytical methods on the idea that
a larger effective backreaction parameter $\tilde{\gamma}$
corresponds to a smaller critical temperature $T_{c}$.
That means the field $\eta(r)$ and $\phi(r)$ coupled to determine the critical temperature.
We argued that negative effective backreaction parameters developing thermodynamically unstable
solutions may be a general properties in various holographic superconductor models.

\begin{acknowledgments}

This work was supported
by the National Natural Science Foundation of China under Grant Nos. 11305097; the
education department of Shaanxi province of China under Grant No. 2013JK0616; the Foundation of Shaaxi University of Technology
of China under Grant No. SLGQD13-23.
This work was also partly finished
during the International Conference
on holographic duality for condensed matter physics
at Kavli Institute for Theoretical Physics China (KITPC), Chinese Academy of
Sciences on July 6-31, 2015.

\end{acknowledgments}

\end{document}